\documentclass[USenglish,twocolumn,colorlinks=true,allcolors=blue]{article}

\ifx\directlua\undefined\ifx\XeTeXcharclass\undefined
  \usepackage[utf8]{inputenc}                           
  \else\RequirePackage[no-math]{fontspec}[2017/03/31]\fi 
  \else\RequirePackage[no-math]{fontspec}[2017/03/31]\fi 
\usepackage[sort&compress,square,numbers]{natbib}
\usepackage[big,online]{dgruyter}

\usepackage{physics}
\usepackage{graphicx}
\usepackage{caption}

\theoremstyle{dgthm}

\theoremstyle{dgdef}

\newcommand{\Baux}{B_{\mathrm{aux}}}
\newcommand{\Jfit}{J_{\mathrm{fit}}}
\newcommand{\gammamod}{\gamma_{\mathrm{mod}}}
\newcommand{\Deltamod}{\Delta_{\mathrm{mod}}}
\newcommand{\tildegammamod}{\tilde{\gamma}_{\mathrm{mod}}}
\newcommand{\tildeDeltamod}{\tilde{\Delta}_{\mathrm{mod}}}

\newcommand{\newtext}[1]{\textcolor{black}{#1}}

\begin{document}

\def\equationautorefname#1#2\null{Eq.#1(#2\null)}
\renewcommand{\figureautorefname}{Fig.}

\articletype{Research Article}

\author[1]{Carlos J. Sánchez Martínez}
\author*[2]{Johannes Feist}
\author*[3]{Francisco J. García-Vidal}
\affil[1]{Departamento de Física Teórica de la Materia Condensada and Condensed Matter Physics Center (IFIMAC), Universidad Autónoma de Madrid, E-28049 Madrid, Spain, E-mail: carlosj.sanchez@uam.es. https://orcid.org/0009-0005-4965-478X}
\affil[2]{Departamento de Física Teórica de la Materia Condensada and Condensed Matter Physics Center (IFIMAC), Universidad Autónoma de Madrid, E-28049 Madrid, Spain, E-mail: johannes.feist@uam.es. https://orcid.org/0000-0002-7972-0646}
\affil[3]{Departamento de Física Teórica de la Materia Condensada and Condensed Matter Physics Center (IFIMAC), Universidad Autónoma de Madrid, E-28049 Madrid, Spain; and Institute of High Performance Computing (IHPC), Singapore 138632, E-mail: fj.garcia@uam.es. https://orcid.org/0000-0003-4354-0982}
\runningauthor{C. J. Sánchez Martínez et al.}
\title{A mixed perturbative-nonperturbative treatment for strong light-matter interactions}
\runningtitle{A mixed perturbative-nonperturbative treatment for strong light-matter interactions}
\abstract{The full information about the interaction between a quantum emitter and an arbitrary electromagnetic environment is encoded in the so-called spectral density. We present an approach for describing such interaction in any coupling regime, providing a Lindblad-like master equation for the emitter dynamics when coupled to a general nanophotonic structure. Our framework is based on the splitting of the spectral density into two terms. On the one hand, a spectral density responsible for the non-Markovian and strong-coupling-based dynamics of the quantum emitter. On the other hand, a residual spectral density including the remaining weak-coupling terms. The former is treated nonperturbatively with a collection of lossy interacting discrete modes whose parameters are determined by a fit to the original spectral density in a frequency region encompassing the quantum emitter transition frequencies. The latter is treated perturbatively under a Markovian approximation. We illustrate the power and validity of our approach through numerical simulations in three different setups, thus offering a variety of scenarios for a full test, including the ultra-strong coupling regime.}
\journalname{Nanophotonics}
\journalyear{2023}
\journalvolume{aop}

\maketitle

\section{Introduction}

In recent years, the development of nanophotonic devices where light is confined
at length scales far below the optical wavelength is leading to new venues for
integrated circuitry, optical quantum computing, solar and medical
technologies~\cite{Koenderink2015}. The nanophotonic structures capable of
obtaining such extreme light confinement are plasmonic (metallic) and hybrid
metallodielectric nanocavities. In particular, the location of a quantum emitter
in close proximity to such nanostructures results in promising enhanced
light-matter interactions, ranging from the enhancement of the spontaneous
emission rate~\cite{Tame2013,Akselrod2014} (known as Purcell
effect~\cite{Purcell1946}) to the possibility of reaching
strong~\cite{Chikkaraddy2016,Baumberg2019,Heintz2021,Li2022} and, even,
ultra-strong light-matter coupling~\cite{Keller2017}.

The complex geometry and the lossy character inherent in these metallic-based
nanodevices define an arbitrary electromagnetic (EM) environment that is open,
dispersive and absorbing. In this scenario, the EM mode spectrum is typically
characterized by arbitrarily broad and overlapping resonances embedded in the
continuum. The quantization of this medium-assisted EM field constitutes a
genuine challenge as losses must be treated explicitly, such that traditional
techniques of quantization fail~\cite{Feist2020}. These difficulties are
formally solved by macroscopic quantum electrodynamics
(QED)~\cite{Huttner1992,Scheel2008}. This framework provides a quantization
scheme for the EM field in arbitrary structures, including dispersive and lossy
materials. The outcome is an EM field described through a four-dimensional
continuum of quantum harmonic oscillators in real space and frequency. Despite
the power and generality of this formalism, \newtext{recently used for exploring the emerging phenomena in nanophotonics~\cite{Hsu2017,Wang2019,Safari2020,Thanopulos2022,Wei2023},} a description based on an extremely
large collection of modes like that represents a clear drawback. On the one
hand, it restricts the direct applicability of this approach to cases where
the EM modes can be treated perturbatively or eliminated by Laplace transform or
similar techniques, and on the other hand, it precludes the desirable application
of standard quantum optics (cavity QED) protocols based on a single or a few
isolated modes interacting with a quantum emitter and capable of accounting for
strong light-matter interactions. Several important steps towards making a
connection with such practical quantum optics approaches have been taken 
in the last decades~\cite{Imamoglu1994, Garraway1997Decay,
Dalton2001}, and quantized few-mode descriptions for specific plasmonic
geometries such as surfaces~\cite{Gonzalez-Tudela2014}, spheres~\cite{Waks2010,
Delga2014, Rousseaux2016, Varguet2019Non-hermitian} or sphere
dimers~\cite{Li2016Transformation, Cuartero-Gonzalez2018} have been obtained.
However, until recently no general frameworks for achieving few-mode field
quantization in arbitrary structures were available, particularly in the case of
hybrid structures where it is necessary to describe modes with quite different
characteristics and mutual coupling.

This was solved in the last few years with two complementary approaches, both
building on the framework of macroscopic QED\@. One relies on using quasinormal
modes~\cite{Ching1998,Kristensen2014}, which are open cavity modes with complex
eigenfrequencies. They are constructed by combining different macroscopic QED
modes, constituting a nonorthogonal basis that is then orthonormalized and
treated with some approximations to arrive at a standard quantum optics
Hamiltonian containing a few lossy discrete modes interacting with each
other~\cite{Franke2019,Franke2022}. The other approach~\cite{Medina2021} is
inspired by tools from the field of open quantum
systems~\cite{Carmichael1993Book,Breuer2007}. It replaces the original EM
environment by a model system involving only a small number of lossy interacting
discrete modes. The model permits the calculation of a compact closed expression
for its spectral density, such that a fitting procedure to the original spectral
density provides a few-mode field quantization of the EM field. In comparison
with the previous quasinormal-mode expansion, this approach requires fewer modes
for convergence. Furthermore, it has recently been extended to the treatment of
both multiple emitters~\cite{Sanchez-Barquilla2022} and the ultra-strong
coupling regime~\cite{Lednev2023}. Its main downside is that, depending on the
complexity of the EM environment, the number of discrete modes required for an
accurate fit can still be larger than ideally wished to ensure low computational
cost.

In the present work, we tackle this issue. We explore an approach capable of reducing the number of modes needed in~\cite{Medina2021} and based on exploiting the underlying physics of the interaction. We divide the spectral density into two contributions in order to separate effectively the part of the EM environment strongly coupled to the emitter from that one which is weakly coupled to it. The strongly coupled environment, \newtext{which induces} non-Markovian \newtext{dynamics}, is treated nonperturbatively using the technique developed in~\cite{Medina2021} of finding an auxiliary few-mode model for such environment. The residual environment is instead treated perturbatively under the assumption of Markovianity, reflecting its effect in an energy shift on the emitter energy levels dubbed Casimir-Polder (CP) energy shift~\cite{Buhmann2012I}. Note that this mixed treatment avoids the use of discrete modes for the part of the environment that is treated perturbatively.

We demonstrate that our model allows the description of the emitter dynamics through a Lindblad-like master equation, even for the ultra-strong coupling regime in which it is well-known that standard Lindblad dissipation terms give rise to unphysical effects. We then test our model validity through numerical calculations of the population of a two-level emitter in the problem of spontaneous emission, in three different setups. The first two exhibit light-matter interactions in the strong coupling regime: one is a canonical test example consisting of a Lorentzian-like spectral density, and the other one is a realistic hybrid metallodielectric nanostructure. The third setup goes beyond exhibiting a real ultra-strong coupling case.

\section{Model}

Our starting point is the general Hamiltonian describing a quantum emitter
linearly coupled to a collection of bosonic modes representing the
medium-assisted EM field:
\begin{equation}\label{eq:discH}
    H = H_e + \sum_{\alpha}\omega_{\alpha}a_{\alpha}^{\dagger}a_{\alpha} + D_e \sum_{\alpha}\left(g_{\alpha}a_{\alpha} + \textup{h.c.} \right),
\end{equation}
where we here and in the following set $\hbar=1$. The emitter is described by
its Hamiltonian $H_e$ and dipole operator $\vec{D}_e = D_e \newtext{d}
\vec{n}$, where all transitions are assumed to be oriented along the same direction $\vec{n}$, and $\newtext{d}$ is a characteristic dipole moment such that
$D_e$ is unitless. The EM modes are described by their annihilation
operators $a_{\alpha}$, frequencies $\omega_{\alpha}$, and coupling to the
emitter $g_{\alpha}$ (which depends on $\vec{n}$ and $\newtext{d}$). The full
information about the light-matter coupling is then encoded in the so-called
spectral density:
\begin{equation}\label{eq:discSP}
    J(\omega) = \sum_{\alpha} \abs{g_{\alpha}}^2 \delta(\omega - \omega_{\alpha}).
\end{equation}
Although our approach is valid for multi-level emitters, from now on we will
consider a two-level system (TLS) with ground state $\ket{g}$, excited state
$\ket{e}$ and transition energy $\omega_{e}$. Under this approximation, the
emitter operators become $H_e = \omega_{e} \sigma^{+}\sigma^{-}$ and
$D_e = \sigma^{+} + \sigma^{-}$, with ladder operators $\sigma^{+}=\ket{e}\bra{g}$
and $\sigma^{-}=\ket{g}\bra{e}$.

We note that the index $\alpha$ in \autoref{eq:discH} and \autoref{eq:discSP} is
a compact notation to represent a set of both discrete and continuous variables
(for which the sum becomes an integral). In particular, within macroscopic QED,
$\alpha$ represents a combined index for 4 continuous (three spatial and one
frequency) and 2 discrete (Cartesian direction and electric or magnetic
excitation) degrees of freedom~\cite{Scheel2008}. The Hamiltonian in
\autoref{eq:discH} then describes the physical system we are interested in: a
quantum emitter interacting with the EM field supported by a nanophotonic
structure, described within the Power-Zienau-Woolley picture~\cite{Power1959} and the
long-wavelength (or dipole) approximation. The spectral density
\autoref{eq:discSP} can then be written in terms of the classical dyadic EM
Green's tensor $\mathbf{G}(\vec{r}, \vec{r}', \omega)$~\cite{Novotny2012, Buhmann2012I}:
\begin{equation}\label{eq:g}
    J(\omega)=\frac{d^2\omega^2}{\pi \varepsilon_0 c^2}\vec{n}\cdot \Im\mathbf{G}(\vec{r}_e, \vec{r}_e, \omega) \cdot \vec{n},
\end{equation}
where $\vec{r}_e$ is the emitter position. \newtext{The Green's tensor of Maxwell's equations~\cite{Scheel2008} fulfills
\begin{equation}\label{eq:pde}
    \left[\nabla \times \frac{1}{\mu(\vec{r},\omega)}\nabla \times - \frac{\omega^2}{c^2}\varepsilon(\vec{r},\omega)\right]\mathbf{G}(\vec{r}, \vec{r}', \omega)=\boldsymbol{\delta}(\vec{r} - \vec{r}'),
\end{equation}
where $\boldsymbol{\delta}\left(\vec{r} - \vec{r}'\right)$ is the Dirac-delta tensor and $\varepsilon(\vec{r},\omega)$ and $\mu(\vec{r},\omega)$ are, respectively, the electric permittivity and magnetic permeability accounting for our electromagnetic configuration. Note that in free-space ($\varepsilon=\mu=1$), the solution of \autoref{eq:pde} is analytical:
\begin{equation}\label{eq:G0}
    \mathbf{G}_0(\vec{r}, \vec{r}', \omega)=\left[\mathbf{I}+\frac{1}{k^2}\nabla \otimes \nabla\right]\frac{e^{ikR}}{4\pi R},
\end{equation}
where $\mathbf{I}$ is the identity tensor, $R=\abs{\vec{r}-\vec{r}'}$ and $k=\omega/c$. In the presence of a nanostructure, the solution of \autoref{eq:pde} is generally no longer analytical but can be written as $\mathbf{G} = \mathbf{G}_0 + \mathbf{G}_s$, where $\mathbf{G}_s$ accounts for the fields scattered by the nanostructure}. Similarly, the spectral density can be split as $J(\omega)=J_0(\omega)+J_s(\omega)$ provided by \autoref{eq:g}:
\begin{subequations}
    \label{eq:J_split}
    \begin{align}
       J_0(\omega) &= \frac{d^2\omega^3}{6\pi^2 \varepsilon_0 c^3}, \label{eq:J0}\\
       J_s(\omega) &= \frac{d^2\omega ^2}{\pi \varepsilon_0 c^2}\vec{n} \cdot \Im\mathbf{G}_s(\vec{r}_e,\vec{r}_e,\omega)\cdot \vec{n}, \label{eq:Js}
    \end{align}
\end{subequations}
where \newtext{in \autoref{eq:J0} we have used that the free-space Green's tensor fulfills} $\vec{n} \cdot \Im\mathbf{G}_0(\vec{r}_e,\vec{r}_e,\omega)\cdot \vec{n} =
\frac{\omega}{6\pi c}$.

\begin{figure}[t!]
\centering
\includegraphics[width=\linewidth]{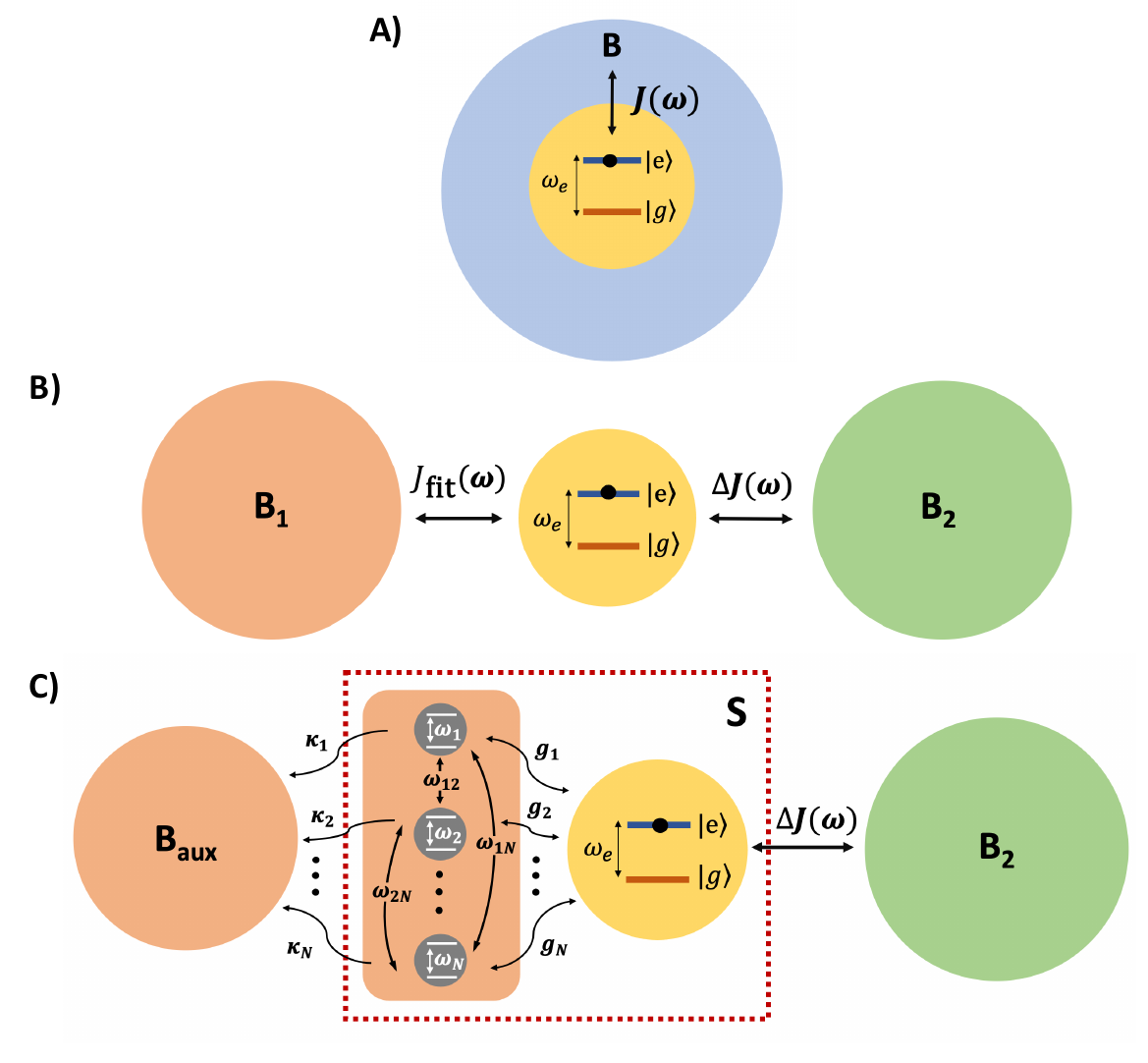}
\captionsetup{justification=justified}
\caption{Graphical representation of our composite system. \textbf{A)} Original configuration: Emitter interacting with a EM bath $B$ through the spectral density $J(\omega)$. \textbf{B)} Original configuration with the EM bath $B$ split into two independent contributions, $B_1$ and $B_2$, to which the emitter coupling is encoded in the spectral densities $\Jfit(\omega)$ and $\Delta J(\omega)$, respectively. \textbf{C)} Model configuration: $B_1$ is substituted by $N$ interacting modes coupled to a spectrally flat auxiliary EM bath $\Baux$. The discrete modes are also coupled to the emitter, conforming a new open quantum system $S$.}
\label{fig:n1}
\end{figure}

The above reflects a more general property: The spectral density can be
rearranged arbitrarily and written as the sum of different contributions that
can be treated independently, with only their sum being physically meaningful.
This can also be understood from \autoref{eq:discSP}, where the sum over modes
$\alpha$ can be obviously split into several sums over arbitrary groups of
indices $\alpha$. We exploit this freedom to write the spectral density as the
sum of two contributions, $J(\omega) = \Jfit(\omega) + \Delta J(\omega)$. The
first, $\Jfit(\omega)$, describes modes close to resonance with the emitter that
can lead to non-Markovian effects such as strong coupling, while the second,
$\Delta J(\omega)$, describes small and/or off-resonant contributions that can
be treated perturbatively. Within this picture, the emitter is coupled to two
independent EM baths $B_1$ and $B_2$ described by $\Jfit(\omega)$ and $\Delta
J(\omega)$, respectively (see \autoref{fig:n1}A-B).

The light-matter interaction described by $\Delta J(\omega)$ can be then
directly treated through a perturbative approach and within the Markov
approximation. This approximation should be valid as long as $\Delta J(\omega)$
is small and flat enough over the bandwidth of frequencies that the emitter is
resonant with. On the contrary, the light-matter interaction with $B_1$,
described by $\Jfit(\omega)$ and characterized by nonperturbative features, is
addressed following the strategy presented in~\cite{Medina2021}: We replace $B_1$ by
an equivalent environment consisting of $N$ interacting discrete modes coupled
to a fully Markovian bath $\Baux$. In the resulting model configuration (see
\hyperref[fig:n1]{Fig.~\ref{fig:n1}C}), the bipartite system $S$
(emitter+discrete modes) is considered as the open quantum system, with
Hamiltonian:
\begin{equation}\label{eq:HS}
    H_S = H_e + \sum_{i,j}^{N} \omega_{ij} a_i^{\dagger}a_j + \sum_{i}^{N} g_i  \left( \sigma^{+} + \sigma^{-} \right)\left(a_i + a_i^{\dagger} \right),
\end{equation}
where $\omega_{ij}$ encodes the mode energies and couplings, $\kappa_i$ their
decay rates, and $g_i$ their coupling to the emitter (including the transition
dipole moment $\newtext{d}$). The value of these parameters is obtained by a nonlinear
fit of $\Jfit$ to the desired region of $J$. This fitting can be performed
relatively straightforwardly as $\Jfit$ is given by a compact expression:
\begin{equation}\label{eq:Jfit}
    \Jfit(\omega) = \frac{1}{\pi} \vec{g} \cdot \Im\left[\frac{1}{\tilde{\mathbf{H}} - \omega}\right] \cdot \vec{g}^T,
\end{equation}
with $\vec{g}=(g_1,g_2,\ldots,g_N)$ and $\tilde{\mathbf{H}}_{ij}=\omega_{ij} - \frac{i}{2}\kappa_i\delta_{ij}$. \newtext{After choosing an initial guess of the parameter values, they are optimized with standard methods of nonlinear fitting to find values that minimize the difference between the physical and the fitted spectral density. Note that the results are not unambiguous, as different sets of parameters can give very similar (or even identical) spectral densities~\cite{Sanchez-Barquilla2021}. Furthermore, the number of modes required to achieve a good fit depends on the complexity of the spectral density and is a manually chosen ``hyperparameter''. While its minimum value is determined by the number of resonances within the fitted window, it can be increased to improve the quality of the fit as required. Note that the number of modes is in this sense not a physically meaningful quantity (only the spectral density is), but a computational parameter that can be chosen to achieve a desired accuracy.}

The coupling of $S$ with the two baths is then treated perturbatively \newtext{following the standard Markovian procedure in open quantum systems textbooks~\cite{Carmichael1993Book,Breuer2007}} (which is exact for $B_\mathrm{aux}$\newtext{~\cite{Tamascelli2018}} and approximate for $B_2$). This leads to a Lindblad master equation for the dynamics of the system $S$:
\begin{multline}\label{eq:merot}
    \frac{d}{dt}\rho_S(t) = -i\left[H_S + H_{CP}, \rho_S(t) \right]\\
    + \gammamod \mathcal{D}_{\sigma^{-}}\left[\rho_S(t) \right]
    + \sum_i^N\kappa_{i} \hspace{0.07cm} \mathcal{D}_{a_i}\left[\rho_S(t) \right],
\end{multline}
where $\mathcal{D}_{o}\left[\rho_S(t) \right] = o \rho_S(t) o^{\dagger} -
\frac{1}{2}\left\lbrace o^{\dagger}o, \rho_S(t) \right\rbrace$ is a standard
Lindblad dissipator, $H_{CP}=-\Deltamod \hspace{0.04cm} \sigma^{+}\sigma^{-}$
encodes the CP energy shift $\Deltamod$, and $\gammamod$ is a decay rate. Both
parameters, $\Deltamod$ and $\gammamod$, arise from the perturbative treatment
of $\Delta J$ within an additional rotating wave approximation for the light-matter
coupling, and are given by
\begin{subequations}
\begin{align}
    \Deltamod &= \mathcal{P} \int_{-\infty}^{\infty}d\omega\frac{\Delta J_s(\omega)}{\omega - \omega_{e}}, \label{eq:delta_mod}\\
    \gammamod &= 2\pi\Delta J(\omega_{e}), \label{eq:gamma_mod}
\end{align}
\end{subequations}
where $\Delta J_s(\omega)=J_s(\omega)-\Jfit(\omega)$ and $\mathcal{P}$ indicates
a principal value integral. Here, $J_s$ instead of the full spectral density
appears as the energy shift, since the free-space contribution $J_0$ leads to a
diverging shift when inserted directly in \autoref{eq:delta_mod}, but gives the
small free-space Lamb shift when treated correctly~\cite{Bethe1947,
Buhmann2012I}. It is thus assumed that its influence is already included in the
emitter transition frequency $\omega_e$. Note also that the integral over
frequencies in \autoref{eq:delta_mod} extends over the full real axis. While
$J(\omega)$ is nonzero only for positive frequencies (at zero temperature, as
considered throughout this manuscript), $\Jfit(\omega)$ is defined and non-zero
for all frequencies.

The approach described above has two potential advantages compared to the one
in~\cite{Medina2021} that it extends: First, it can be used to reduce the number
of auxiliary coupled oscillators that have to be included in $S$ by only fitting
a reduced part of the spectrum, and second, it can be used to mitigate any
inaccuracies in the fit by including the resulting correction $\Delta J(\omega)
= J(\omega) - \Jfit(\omega)$ in the master equation, albeit only within the
Markovian approximation. Below, we investigate the accuracy of the resulting
model in different scenarios. It can be anticipated that the model will work
well if those parts of the spectral density that induce non-Markovian dynamics
on the emitter are well-described by the auxiliary model described by
$\Jfit(\omega)$, which usually requires that the fit is accurate at least within
a spectral window close to the emitter transition frequency. \newtext{We note
that if the whole light-matter interaction is in the weak-coupling regime, it
can be treated fully perturbatively, and the model is not needed. This case is
equivalent to setting $\Jfit(\omega) = 0$.}

\newtext{For the case of interest where the overall coupling is non-Markovian, a
criterion to estimate the validity of the splitting can be formulated by
utilizing that the enforced good agreement between $J(\omega)$ and
$\Jfit(\omega)$ in the spectral region close to the emitter frequency implies
that $\Delta J(\omega)$ is small within that region (and presumably larger
outside), so that it naturally splits into the two regions $\omega<\omega_e$ and
$\omega>\omega_e$. The validity of the perturbative treatment of $\Delta
J(\omega)$ can then be checked by treating each of the two regions separately
and veryifying that the resulting interaction is indeed perturbative. This can
be done by, for example, checking that the ``reaction
mode''~\cite{Martinazzo2011Communication, Woods2014, Sanchez-Barquilla2021} that
subsumes the bath-emitter interaction in each region is perturbatively coupled
to the emitter. This leads to the condition $\beta_\pm =
\frac{g_\pm^2}{(\omega_\pm - \omega_e)^2} \ll 1$. Here, the effective coupling
is given by $g_\pm^2 = \int_{A_\pm} \Delta J(\omega) \mathrm{d}\omega$ and the
effective frequency by $\omega_\pm = \int_{A_\pm} \omega \Delta J(\omega)
\mathrm{d}\omega / \int_{A_\pm} \Delta J(\omega) \mathrm{d}\omega$, where $A_+$
($A_-$) is the region of frequencies where $\omega>\omega_e$
($\omega<\omega_e$). In the
examples shown below, the condition $\beta_\pm \ll 1$ is always fulfilled.}

The above definition of $\Delta J(\omega)$ as the difference between the
physical and the fitted spectral density unveils a subtle point: while both
$J(\omega)$ and $\Jfit(\omega)$ correspond to the spectral density of physical
systems and are thus strictly non-negative functions, $\Delta J(\omega)$ does
not necessarily fulfill this constraint. This does not present a particular
problem for the CP-shift $\Deltamod$, which in any case can be a positive or
negative energy shift, but can appear problematic for the decay rate
$\gammamod$, since the Lindblad master equation is not a completely positive map
if $\gammamod<0$, and the resulting terms do not describe decay, but
``anti-decay'', i.e., an exponential growth of population.\footnote{Note that is
not the same as a ``pumping'' Lindblad term, which corresponds to a normal
Lindblad term with positive rate and an associated operator that lifts the
system to a state with higher energy (e.g., $\Gamma_\mathrm{pump}
\mathcal{D}_{\sigma^+}[\rho]$).} \newtext{Note that in principle, the derivation
of the Lindblad master equation requires that the spectral density be positive,
and allowing for negative rates is thus not strictly justified. In this sense,
anti-decay terms are a generalization of existing results to a regime outside
their original range of validity. In the context of open quantum systems,
similar generalizations are commonly done with negative-frequency harmonic
oscillators, which are not eigenstates of a physical potential, but can be
useful tools to generalize approaches to new regimes~\cite{Tamascelli2019}.}

We will show below that the appearance of negative rates is not an issue in
practice when the description is sufficiently accurate. This is consistent with
similar results found for the Bloch-Redfield approach~\cite{Eastham2016}, i.e.,
a perturbative treatment of a bath within the Born-Markov approximation, which
can induce negative decay rates if no additional secular approximation is
performed~\cite{Breuer2007}. For most cases we study, the problem does not
appear, since the spectral density is fitted accurately close to the emitter
frequency and thus $\gammamod = 2\pi \Delta J(\omega_e) \approx 0$. However, in
\autoref{sec:ultrastrong} we treat a system in the ultrastrong-coupling regime
where counterrotating terms in the light-matter interaction cannot be neglected.
For this system, we show that a negative-rate ``anti-Lindblad'' term can
efficiently cancel unphysical artificial pumping effects that otherwise
appear~\cite{Lednev2023}. \newtext{This term arises naturally from the
perturbative treatment of $\Delta J$} when\newtext{, unlike \autoref{eq:merot},}
the rotating-wave approximation in the light-matter coupling is not performed,
such that the resulting generalized Lindblad-like master equation is
\begin{multline}\label{eq:meUSC}
    \frac{d}{dt}\rho_S(t) = -i\left[H_S + H_{CP} + \tilde{H}_{CP}, \rho_S(t) \right] \\
    + \gammamod \mathcal{D}_{\sigma^{-}} \left[\rho_S(t) \right] + \tildegammamod \mathcal{D}_{\sigma^{+}}\left[\rho_S(t) \right] \\
    + \sum_i^N\kappa_{i} \mathcal{D}_{a_i}\left[\rho_S(t) \right],
\end{multline}
which contains both an extra CP term $\tilde{H}_{CP} = -\tildeDeltamod
\sigma^{+}\sigma^{-}$, where
\begin{equation}\label{eq:delta_mod_tilde}
    \tildeDeltamod = - \mathcal{P} \int_{-\infty}^{\infty}d\omega \hspace{0.04cm} \frac{\Delta J_s(\omega)}{\omega + \omega_{e}},
\end{equation}
and \newtext{the additional Lindblad term with rate $\tildegammamod = 2\pi
\Delta J(-\omega_e)$}. This rate is always negative since $J(\omega) = 0$ for
negative frequencies $\omega<0$, while $\Jfit(\omega)\geq 0$ for any $\omega$,
\newtext{such that the term becomes an ``anti-Lindblad'' one as described above.
Observe as well that \autoref{eq:delta_mod_tilde} is identical to
\autoref{eq:delta_mod} performing the substitution $\omega_e \rightarrow
-\omega_e$ (the overall minus sign is a matter of convention to write both CP
energy terms preserving the same form)}.

\begin{figure*}[!ht]
\centering
\includegraphics[width=\linewidth]{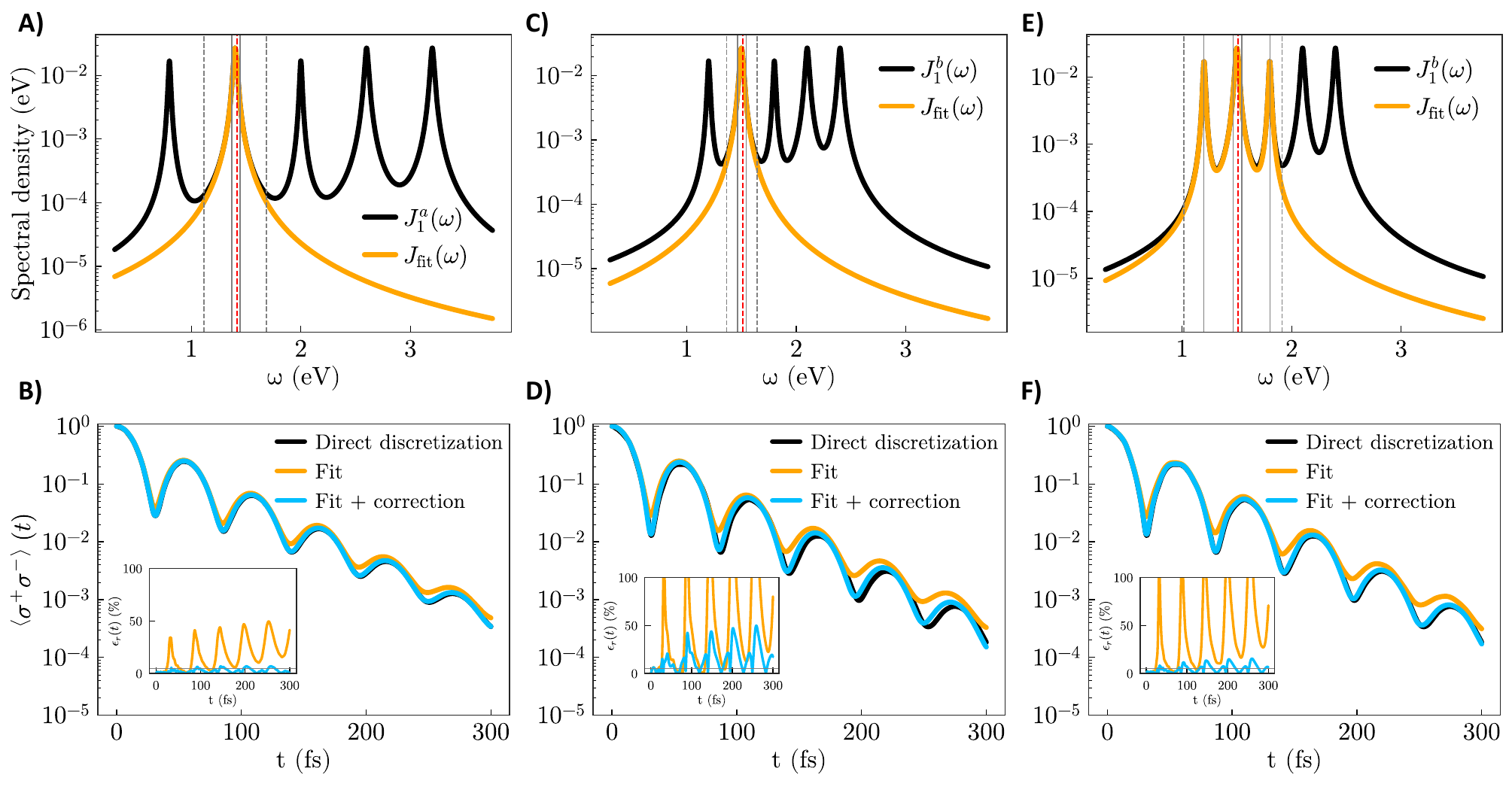}
\captionsetup{justification=justified}
\caption{Numerical simulations on the test Lorentzian-like spectral density. \textbf{Left column (A-B):} Case with well-separated resonances. \textbf{A)} Spectral density splitting with a $1$-mode model for $\Jfit$. The residual spectral density is not shown. The emitter transition frequency is indicated with a dashed red line and the fit range is comprised between the two dashed gray lines. The solid gray lines indicate the energy positions of the eigenstates of $H_S$. \textbf{B)} Emitter excited-state population calculated from $3$ different approaches: direct discretization (black), only considering $\Jfit$ (orange) and our full model (blue). The inset shows the relative error with respect to the exact result, where a solid gray horizontal line is traced at $5\%$. \textbf{Central column (C-D):} The same as left column but for a squeezed spectral density. \textbf{Right column (E-F):} The same as central column but with a $3$-mode model for $\Jfit$. Parameters: (A-B) $\omega_e=1.4155$, $\Deltamod=0.0021$, \newtext{$\gammamod\approx 0$} [eV]; (C-D) $\omega_e=1.5135$, $\Deltamod=0.0043$, \newtext{$\gammamod\approx 0$} [eV]; (E-F) $\omega_e=1.5135$, $\Deltamod=0.0041$, \newtext{$\gammamod\approx 0$} [eV].}
\label{fig:n2}
\end{figure*}

\section{Results}

We test the accuracy and regime of validity of our model by performing numerical
simulations of the excited-state population of a TLS, $\left\langle
\sigma^{+}\sigma^{-} \right\rangle (t)$, for the paradigmatic problem of
spontaneous emission. \newtext{Notice that our model allows the computation of
expectation values of any observable $O$, $\left\langle O \right\rangle(t) =
\operatorname{Tr}\{ O\rho_S(t) \}$, since it provides the density matrix
operator $\rho_S(t)$. Furthermore, it is not restricted to the single-excitation
subspace~\cite{Sanchez-Barquilla2022,Lednev2023}}.

We consider three different model EM environments: the first
corresponds to a simple model where the spectral density is described by a sum
of Lorentzian resonances, the second one is a realistic hybrid metallodielectric
nanostructure, and the third one is a single-mode setup corresponding to a
two-level emitter under ultrastrong coupling to a single physical mode. In all
three systems, the light-matter coupling is strong enough to obtain
non-Markovian effects, as the weak (Markovian) coupling regime can already be
described accurately by fully perturbative approaches (equivalent to setting
$\Jfit(\omega)=0$ in our model). The first two systems are treated within the
rotating-wave approximation and described by \autoref{eq:merot}, while the third
one is within the ultra-strong coupling regime where this approximation is not
valid and the effective master equation is given by \autoref{eq:meUSC}. In all
cases, we compare the results obtained with our approach with an exact solution
obtained by direct discretization of the original Hamiltonian in
\autoref{eq:discH} (which is numerically feasible for propagation over short
times and when no decoherence apart from that induced by the bath is present,
such that the dynamics is purely coherent).

\subsection{Lorentzian model spectral density}

We start with a test case consisting of an example EM environment characterized
by a spectral density that is the sum of Lorentzian resonances,
\newtext{$J_{1}(\omega) = \sum_i \frac{g_i^2}{\pi} \frac{\kappa_i/2}{(\omega -
\omega_i)^2 + (\kappa_i/2)^2}$}, which corresponds to the non-interacting limit
of \autoref{eq:Jfit} ($\omega_{ij}=0$ for $i\neq j$). The non-interacting
character of the modes and the flexibility for tuning their strength $g_i$,
width $\kappa_i$, and frequencies $\omega_i$ offer an ideal scenario for gaining
intuition on the model. In particular, since the form is exactly that of
$\Jfit(\omega)$, the splitting into $\Jfit(\omega)$ and $\Delta J(\omega)$ can
be performed by just including some of the sum terms in $\Jfit$ without the need
to perform any fitting. We use a $5$-mode spectral density, considering two
different situations: In the first case, $J_1^a(\omega)$, the $5$ peaks are
spectrally well-separated, with a regular spacing of $\omega_{i+1} - \omega_{i}
= 0.6$~eV, while in the second case, $J_1^b(\omega)$, the separation between the
peaks is reduced by half. In each configuration we study the scenario where the
five modes fulfill the condition $g_i/\kappa_i>1$, guaranteeing the strong
coupling regime (as will below reflected in a reversible dynamics). Note that
this regime leads to the formation of hybrid light-matter states called
polaritons (eigenstates of $H_S$) whose frequencies determine the actual
dynamics, and which are, in general, different from the frequencies in the
uncoupled Hamiltonian.

First, we focus on the results for the configuration with well-separated
resonances, presented in \hyperref[fig:n2]{Fig.\ \ref{fig:n2}A-B}, with the
emitter resonant with the second mode. We use the simplest choice for $\Jfit$: a
$1$-mode model, treating nonperturbatively only the closest resonance to the
emitter transition frequency. The rest of the spectral density is treated
perturbatively through $\Delta J$. The splitting of the spectral density is
displayed in \hyperref[fig:n2]{Fig.\ \ref{fig:n2}A} indicating also the emitter
transition frequency (dashed red line), the range over which the fitted spectral
density is accurate (dashed gray lines), and the energies of the formed
polaritons (solid gray lines). The results of time propagation (see
\hyperref[fig:n2]{Fig.\ \ref{fig:n2}B}) show clearly that the dynamics is
produced much more accurately by our model (blue lines) than by the use of only
$\Jfit$ while ignoring the perturbative correction due to $\Delta J(\omega)$
(orange lines). This is corroborated by the inset, which shows that the relative
error, $\epsilon_r(t) = |\langle\sigma^+\sigma^-\rangle -
\langle\sigma^+\sigma^-\rangle_\mathrm{exact}| /
\langle\sigma^+\sigma^-\rangle_\mathrm{exact}$, stays below about $5\%$ for the
entire dynamics, while it reaches $50\%$ when only $\Jfit$ is used.

The results of reducing the spacing between resonances (spectral density
$J_1^b(\omega)$) are presented in \hyperref[fig:n2]{Fig.\ \ref{fig:n2}C-D}.
Using a single-mode model as in the previous configuration now presents much
larger deviations from the exact results. The reason for this is that the
energies of the two polaritons formed by strong coupling between the emitter and
the resonant mode are now much closer to the two nearest-non-fitted resonances,
such that these two resonances also influence the emitter dynamics in a
nonperturbative way that cannot be reflected in the CP energy shift, leading to
multimode strong coupling effects. Including the two closest additional
resonances in $\Jfit$ leads to a $3$-mode model, with results displayed in
\hyperref[fig:n2]{Fig.\ \ref{fig:n2}E-F}. As could be expected, our approach now
again works very well, highlighting the necessity of including a sufficiently
wide frequency range in the fitted spectral density.

This first test example thus provides significant insight on the mixed
nonperturbative-perturbative approach. First, we can deduce that the main
requisite for its success is the inclusion in $\Jfit$ of all the spectral
density contributions that lead to non-Markovian and strong-coupling effects and
that cannot be captured accurately through a perturbative procedure. Second,
even if this identification and fit is performed accurately, the role of the
perturbative energy shift is fundamental to achieve an accurate description when
the spectral density is non-negligible outside the fitted region. This
demonstrates that the original goal can indeed be achieved: the number of
discrete modes that must be included in $\Jfit$ to obtain accurate results can
be significantly reduced compared to the case where $\Jfit$ is used for
describing the whole spectral density.

\subsection{Realistic nanostructure}

\begin{figure}[!h]
\centering
\includegraphics[width=\linewidth]{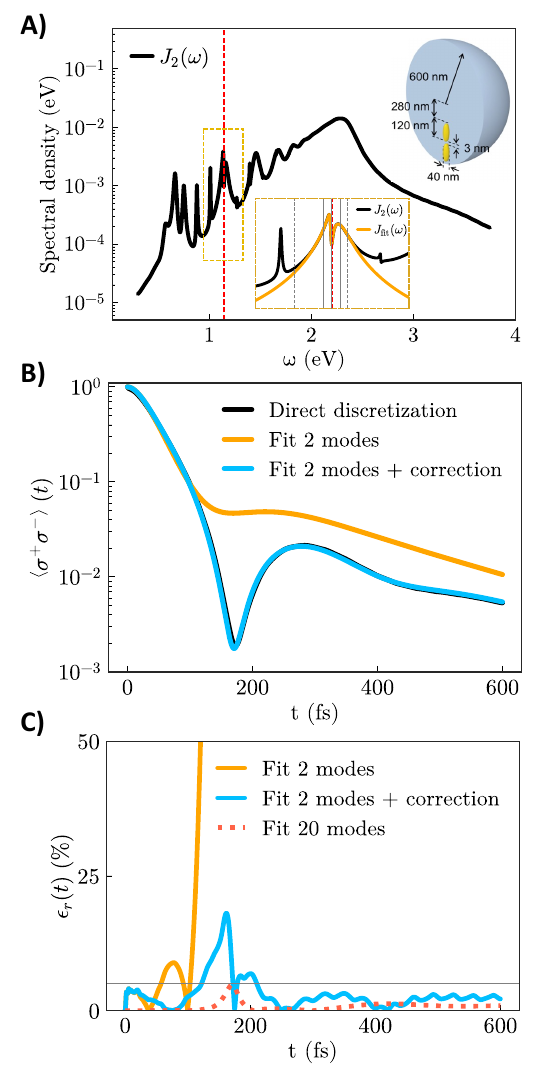}
\captionsetup{justification=justified}
\caption{Numerical simulations on the hybrid metallodielectric nanostructure. \textbf{A)} Spectral density of the system. \newtext{The upper right inset displays a sketch of the metallodielectric nanostructure}. The \newtext{bottom} inset zooms in the $2$-mode spectral density splitting in the region close to the emitter transition frequency. \textbf{B,C)} The same as the second row in \hyperref[fig:n2]{Fig. \ref{fig:n2}} but displaying the relative error in a separated figure (C). The dotted red line indicates the relative error obtained in~\cite{Medina2021} reproducing the whole spectral density with a $20$-mode fit. Parameters: $\omega_e=1.1445$, $\Deltamod=0.0093$, \newtext{$\gammamod\approx 0$} [eV].}
\label{fig:n3}
\end{figure}

These notions are confirmed with the study of the same realistic hybrid
metallodielectric nanostructure treated in~\cite{Medina2021}. It consists of a
dielectric GaP microsphere of radius $600$~nm embedding two $120$~nm long silver
nanorods separated by a $3$ nm gap and substantially displaced from the center
of the sphere \newtext{(see the upper right inset in \hyperref[fig:n3]{Fig.\ \ref{fig:n3}A})}. The emitter is located in the center of the gap, with parameters chosen to represent InAs/InGaAs quantum dots~\cite{Eliseev2000}, with
transition energy $\omega_{e} = 1.1445$~eV and transition dipole moment $\newtext{d} =
0.55$~e\,nm. The hybrid nature of this structure results in a more complex
spectral density $J_2(\omega)$. It is characterized by Fano-like profiles that
indicate interference effects between the different modes supported by the
microsphere and the nanorods (see \hyperref[fig:n3]{Fig.\ \ref{fig:n3}A}).

This complex spectral density is represented in~\cite{Medina2021} in a fully
nonperturbative approach through a fit using $20$ interacting modes. To
illustrate the power of our approach, we use a $2$-mode model for $\Jfit$,
including only the two interacting modes close to resonance with the emitter
\newtext{(see the bottom inset in \hyperref[fig:n3]{Fig.\ \ref{fig:n3}A})}. As shown in
\hyperref[fig:n3]{Fig.\ \ref{fig:n3}B-C}, this is enough to obtain a reliable
description of the emitter dynamics, with a relative error typically on the
few-percent level. We note that the large maximum observed in the relative error
close to $t = 200$~fs is a consequence of the small value of the population at
that point.

The choice of this minimal model for $\Jfit$ is inspired by the fact that,
although the light-matter interaction in this setup is strong (see the clear
reversible behavior close to $t=200$~fs), most of the modes do not in fact enter
the strong coupling with the emitter, but instead only contribute an additional
effective energy shift. This example is thus a clear demonstration of the power
of our model when the frequencies that are on resonance with the emitter are
correctly identified and the perturbative procedure can be safely performed,
resulting in a large reduction of the number of discrete modes required.

\subsection{Extension to the ultra-strong coupling regime}\label{sec:ultrastrong}

We finally illustrate the power of our model in the ultra-strong coupling
regime. We study the same setup analyzed in~\cite{Lednev2023}. It consists of a
physically allowed extension of the quantum Rabi model, and is described by a
spectral density corresponding to a single harmonic oscillator with frequency
$\omega_c$ coupled to an Ohmic ``background'' bath. The resulting spectral
density can be written as:
\begin{equation}
    J_3(\omega) = \theta(\omega)\frac{2g^2}{\kappa}\frac{\kappa \omega_c \omega}{\left(\omega^2 - \omega_c^2 \right)^2 + \kappa^2\omega^2},
\end{equation}
where we use the same parameters as in~\cite{Lednev2023}:
$\omega_c=\omega_e=0.58$~meV, $g=0.25$~meV and $\kappa=0.1$~meV, which are
typical for Landau polaritons formed in semiconductor quantum wells in the USC
regime~\cite{Paravicini-Bagliani2019, Forn-Diaz2019, Keller2020}. Here, $g$ represents the
coupling between the emitter and the mode, and $\kappa$ the losses of the mode.
We note that this spectral density, as any physical spectral density, is
non-zero only for positive frequencies.

\begin{figure}[!h]
\centering
\includegraphics[width=\columnwidth]{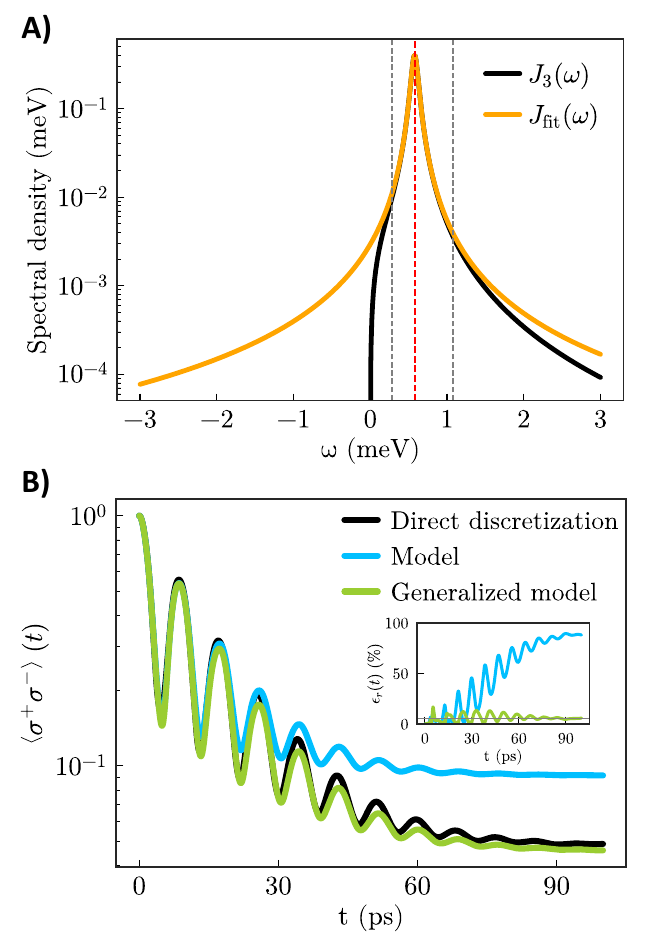}
\captionsetup{justification=justified}
\caption{Numerical simulations on the single-mode setup supporting ultra-strong coupling effects. \textbf{A)} Spectral density splitting with a $1$-mode model for $\Jfit$. \textbf{B)} Emitter population: the blue line is computed through \autoref{eq:merot}, while the green line results from the generalized \autoref{eq:meUSC}. The absolute relative error associated with the two models is displayed in the inset. Parameters: $\omega_e=0.58$, $\Deltamod=0.0026$, $\tildeDeltamod=0.0026$, \newtext{$\gammamod\approx 0$},  $\tildegammamod=-0.0046$ [meV].}
\label{fig:n4}
\end{figure}

The results obtained using a $1$-mode $\Jfit$ are presented in
\hyperref[fig:n4]{Fig.\ \ref{fig:n4}A-B}. Note that $\Jfit$ extends to negative
frequencies, see \hyperref[fig:n4]{Fig.\ \ref{fig:n4}A}. This cannot be avoided
for a single-mode fit. While the flexibility of the coupled-oscillator model can
be exploited to suppress these negative-frequency contributions~\cite{Lednev2023},
this requires the use of several additional auxiliary oscillators. When only a
single mode is used for the fit and no perturbative corrections are performed,
the presence of the negative-frequency components leads to artificial pumping
effects, resulting in an unphysically large emitter population at later times.
This is reflected in the results of \hyperref[fig:n4]{Fig.\ \ref{fig:n4}B},
where the blue line shows the results obtained with \autoref{eq:merot}, with a
severely overestimated population, in particular in the steady state (reached at
around $90$~ps). However, when the perturbative corrections are included as
described in the theory section, \autoref{eq:meUSC}, the presence of the
``anti-Lindblad'' term with associated negative rate $\tildegammamod$ cancels
the unphysical pumping effects, and the results (green line) are much closer to
the exact ones (black line) at essentially the same numerical cost as the
single-mode model. As the inset shows, the relative error within this method
stays low for the whole dynamics, and the steady-state population is reproduced
reasonably well.

\section{Conclusions}

We have presented an approach for describing light-matter interactions in
arbitrarily complex nanophotonic systems in any coupling regime by using a mixed
nonperturbative-perturbative description \newtext{extending} our previously developed
few-mode quantization~\cite{Medina2021}. The approach is based on a splitting of the
spectral density, $J(\omega)$, in order to effectively separate the part
responsible for the non-Markovian and strong-coupling-based emitter dynamics,
$\Jfit(\omega)$, from that which can be treated as a perturbation, $\Delta
J(\omega)$. The former is represented by a minimal collection of lossy
interacting discrete modes coupled to fully Markovian background baths, while
the latter is treated perturbatively with standard open quantum systems theory,
leading to an energy shift on the emitter energy levels and additional Lindblad
dissipator terms (which can contain negative dissipation rates). All this
information is encoded in a compact simple Lindblad master equation.

We have tested our methods by calculating the population dynamics of an
initially excited TLS in three different EM environments of varying complexity,
investigating the strong and ultra-strong coupling regimes. We find that our
model works accurately as long as $\Jfit$ is accurate over a sufficiently large
frequency range to capture all non-Markovian effects. This condition can be
fulfilled by identifying the spectral density region directly coupled to the
relevant transitions frequencies of the system. The remaining spectral density
can then be safely treated perturbatively. As a result, the final model achieves
an accurate description with a significantly reduced numerical cost compared to
the full model fitting the spectral density over its whole bandwidth.
		
\begin{acknowledgement}
We thank Maksim Lednev for fruitful discussions.
\end{acknowledgement}

\begin{funding}
This work was supported by the Spanish Ministry for Science and
Innovation-Agencia Estatal de Investigación (AEI) through the FPI contract No. PRE2022-101819 as well as grants PID2021-125894NB-I00 and CEX2018-000805-M (through the María de Maeztu program for Units of Excellence in R\&D).
\end{funding}

\begin{authorcontributions}
All the authors have accepted responsibility for the entire content of this submitted manuscript and approved submission.
\end{authorcontributions}

\begin{conflictofinterest}
The authors declare no conflicts of interest regarding this article.
\end{conflictofinterest}

\bibliographystyle{apsrev4-2}
\nocite{apsrev41Control}
\bibliography{apsrevbibopts,references}

\end{document}